\documentclass[runningheads]{llncs}
\usepackage{graphicx}
\usepackage{eurosym}
\usepackage[verbose]{placeins}
\usepackage{amsmath}
\usepackage{amssymb}
\usepackage{array}
\usepackage[
    layout = inline,
    final]{fixme}                  	
\fxsetup{
    status=final,                               
    layout=inline,                              
    theme=colorsig                              
}
\FXRegisterAuthor{ck}{ack}{CK}
\usepackage[chapter]{algorithm}
\usepackage{algorithmicx}
\usepackage[]{algpseudocode}

\DeclareMathOperator*{\argmax}{arg\,max}

\begin{document}
\title{Charging control of electric vehicles using contextual bandits considering the electrical distribution grid}
%
%
\author{Christian~R{\"o}mer\inst{1}\orcidID{0000-0003-4386-9323} \and
Johannes~Hiry\inst{2}\orcidID{0000-0002-1447-0607} \and
Chris~Kittl\inst{2}\orcidID{0000-0002-1187-0568} \and
Thomas~Liebig\inst{1}\orcidID{0000-0002-9841-1101} \and
Christian~Rehtanz\inst{2}\orcidID{0000-0002-8134-6841}
}
\authorrunning{C. R{\"o}mer et al.}
%
\institute{{TU Dortmund University\\Department of Computer Science\\Otto-Hahn-Str. 12, 44227 Dortmund, Germany\\\email{\{christian.roemer, thomas.liebig\}}@cs.tu-dortmund.de}\\\url{http://www.cs.tu-dortmund.de}
\and
TU Dortmund University\\Institute of Energy Systems, Energy Efficiency and Energy Economics\\Emil-Figge-Str. 70, 44227 Dortmund, Germany\\\email{\{johannes.hiry, chris.kittl, christian.rehtanz\}}@tu-dortmund.de
\url{http://www.ie3.tu-dortmund.de/}}
\maketitle              
\begin{abstract}
With the proliferation of electric vehicles, the electrical distribution grids
are more prone to overloads. In this paper, we study an intelligent pricing and power control mechanism based on contextual bandits to provide incentives for distributing
charging load and preventing network failure. The presented work combines the
microscopic mobility simulator SUMO with electric network simulator SIMONA and
thus produces reliable electrical distribution load values. Our experiments
are carefully conducted under realistic conditions and reveal that conditional
bandit learning outperforms context-free reinforcement learning algorithms and
our approach is suitable for the given problem.
As reinforcement learning algorithms can be adapted rapidly to include new information we assume these to be suitable as part of a holistic traffic control scenario.

\end{abstract}
\section{Motivation}
The amount of street based individual traffic is rising world wide.
In Germany, the amount of registered passenger cars has risen by 20.9\% between the years 2000 and 2018~\cite{destatisVerkehrAktuell2018}.
At the same time use of electric vehicles is getting more popular: Since the year 2006 the number of registered vehicles in Germany has risen from 1931 to 53861~\cite{statistaElektroautos2018}.
This can be seen as an indicator for a world wide development.
When such a great number of vehicles enter the market it has to be considered that these act as new electric loads in the electrical grid.
This additional load needs to be included by the grid operators in their future planing.
According to Uhlig et al.~\cite{uhligEmobilityIntegration2014} the charging of
electric vehicles (EV) in low voltage grids occurs mainly in the early to late evening, where load peaks are already common.
The combination of already existing peaks and the additional load due to the charging of EVs can lead to critical loads in the involved operating resources.
Schierhorn and Martensen~\cite{schierhornBedeutungElektromobilitaet2016}~found that first line overloads occur at EV-market coverages of 8\%, when vehicles are allowed to charge at all times without restrictions or sophisticated charging strategies.
In this paper we investigate the load reduction by a smart charging strategy.

The situation gets more complicated by the emergence of renewable energies and the liberalization of the energy market, leading to challenging daily fluctuations with partly opposing targets.
While grid operators need to concern themselves to uphold a high service quality by keeping the additional strain on the grid caused by electric vehicles in reasonable boundaries the charging station operators try to maximize utilizations on their assets.

\section{Related Works}
Waraich et al.~\cite{waraichInvestigations2013} have examined an approach for researching the effects of electric vehicles on the power grid and possible strategies for smart charging and prevention of power shortages.
They integrated the simulation tool \textsc{MATSim} for the simulation of traffic flows and energy usages with a ''PEV management and power system simulation (PMPSS)'', which simulates power electrical grids and EV-charging stations.
In the simulation they equipped each subnet with a ''PEV management device'', which optimizes the charging of electric vehicles.
The optimization routine changes a price signal, which depends on the grid's state, the number of charging vehicles and their urgency.
The vehicles react to the price signal within the scope of the MATSim simulation cycle.
The authors survey different scenarios and the effects of the EV-charges to the power grid regarding average and peak loads over the course of a day.
The management device knows the exact daily routine of every vehicle and can use that to plan ahead.

In \cite{sundstrom2010optimization}, the authors present an optimization method
to plan charging of electric vehicles, main focus of their study is the
decision on a charging time interval to keep performance of the individual EV
and reduce energy costs.
In contrast, our work will adjust either energy prices or charging power such that the quality of service is guaranteed by the smart grid.
An interesting related study was published in \cite{vagropoulos2013optimal} and discusses bidding strategies of electric vehicle aggregators on the energy market.

The application of machine learning methods, especially reinforcement learning, to problems in planning and operation of power grids does not seem to be well researched yet.
Vlachogiannis et al.~\cite{vlachogiannisReactivePowerControl2004} used a Q-learning algorithm for reactive power control, finding that while the used algorithm was able to solve the problem it took a long training phase to converge.
They however considered neither EVs nor renewable (weather-dependent) energy generation.
Other authors used learning methods to provide a frequency regulation service via a vehicle-to-grid market~\cite{wuFrequencyRegulation2012} or optimize a residential demand response of single households using an EV-battery as a buffer for high demand situations~\cite{oneillResidentialDemandReponse2010}.

The usage of bandit algorithms and the test environment was motivated by the successful application to the problem of avoiding traffic jams using a self-organizing trip-planning solution in \cite{sotzny17}.

\section{Fundamentals}
This section describes the fundamentals regarding electric power grids, electric mobility and the used learning algorithms needed for comprehension of this work.

\subsection{Electric power grids}
An electric power grid encompasses all utilities required for the transmission and distribution of electric power, like cables, overhead lines, substations and switching devices.
Generally the operational cost of utilities are higher for higher voltages.
However, transmitting great powers over great distances is only possible with high voltages due to transmission losses.
For obtaining the maximal cost effectiveness, the system is segmented into hierarchical levels with each level having a specific purpose.
Extra high and high voltage grids transmit power over large distances and connect large scale industry, cities and larger service areas with great and middle sized power plants.
The purpose of medium voltage grids, which are fed by the high voltage grids and smaller plants, is to supply industry, large scale commercial areas, public facilities and city or land districts.
Low voltage grids make out the ''last mile'' to supply residential or commercial areas and are fed by the medium voltage grids or very small plants like personal photovoltaic systems.


\subsection{Electric mobility}
The propulsion technologies in the electric mobility can be roughly divided into three categories: a) hybrid vehicles, which use two different energy sources for propulsion. Most of the time these are one gasoline engine and one electric engine b) plug-in hybrid vehicles, which are hybrid vehicles that can be connected to the power grid, enabling it to charge the battery while parking c) battery electric vehicles, which do not have a gasoline engine (expect sometimes a so called \emph{range extender}, which however is not directly connected to the powertrain).
In this work we will concentrate on the pure battery electric vehicles (BEV).

Various charging technologies exist for connection BEVs to the electrical power grid. The specifications for conductive charging technologies are mainly defined in the IEC\,61851. An international as well as an european and german norm for inductive charging is currently under development (DIN\,EN\,61980 based on IEC\,61980 and others). Due to the lack of a frequent use of inductive charging technologies they are not consindered further in this work. Hence, considering only conductive charging, one main distinction can be made between alternate current (AC) and direct current (DC) charging. AC charging can be further subdivided depending on the needed maximum power, number of used phases as well as the grid coupler. 

\newcolumntype{M}{>{\centering\arraybackslash}m{\dimexpr.25\linewidth-2\tabcolsep}}
\renewcommand{\arraystretch}{1.2}
\begin{table}
	\caption{Charging modes defined by the DIN-EN-61851}\label{tab1}
	\label{table:1}
	\begin{tabular}{ l  m{\dimexpr.45\linewidth-2\tabcolsep}  l l }
		\hline
		\bfseries Mode & \bfseries Definition & \bfseries Technology & \bfseries Communication \\ 
		\hline
		 1 & Direct household socket connection  & AC, 1 or 3 phase(s) & none\\
		\hline
		 2 & same as 1 plus in-cable control and protective device (IC-CPD)/low level control pilot function & same as 1 & Control Pilot\\\hline
		 3 & Dedicated charging station "wallbox" & AC, 1 or 3 phase(s) & Control Pilot\\ \hline
		 3 & Dedicated charging station & AC, 1 or 3 phase(s) & Powerline (PLC)\\ \hline
		 4 & Dedicated charging station & DC & Powerline (PLC)\\
		\hline
	\end{tabular}
\end{table}

Depending on the kind of charging infrastructure available, one can distinguish between uncontrolled and controlled charging. Uncontrolled in this context means, that the maximum installed power of the charging station is available to the connected car for the whole charging process. During the process there are no external interventions by other entities of the electrical power system (e.\,g. distribution grid operator (DSO)) nor any load shifting or charging strategies executed. This kind of charging can be provided by any of the charging modes shown in Table \ref{table:1}.
In the controlled case, the available installed power can be altered within the technical limits. Specifically, controlled charging can be used to reduce the load on the electrical grid by shifting the charging process from times with high overall grid utilization to times with a lower grid utilization or to carry specific charging strategies for an electric vehicle fleet. 
This process can either be carried out in a centralized (e.\,g. the DSO executes load curtailment actions) or a decentralized (e.\,g. the charging station reduces its power by itself) way. The centralized approach is only possible if the necessary communication infrastructure is available. Hence, only charging modes 3 with PLC or 4 are suitable for the centraliced controlled charging. 

\subsection{LinUCB algorithm for contextual bandits}
\begin{algorithm}[t]
	\caption{LinUCB according to Li, Chu, Langford und Schapire~\cite{liContextualBandits2010}.}
	\label{linucbdisjoint}
	\begin{algorithmic}[1]
		\State Input: $\alpha \in \mathbb{R}_+$, Context dimension $d$, Arms $A$
		\ForAll{arm $a \in A$}
		\State Initialize context histories $A_a$ and reward histories $b_a$
		\State [Hybrid] Initialize shared context history $B_a$ and shared reward history $b_0$
		\EndFor
		\ForAll{round $t = 1,2,3,\dots,T$}
		\State Observe context for all arms $a \in A_t$: $x_{t,a} \in \mathbb{R}^d$
		\ForAll{Arm $a \in A_t$}
		\State Using the context and reward history $A_a$ and $b_a$ do a ridge regression,\\\hspace{1cm}updating the coefficients $\hat{\theta}_a$. Using the coefficients $\hat{\theta}_a$ and the current context\\\hspace{1cm}vector $x_{t,a}$ determine the expected reward $p_{t,a}$.
		\State [Hybrid] Besides $\hat{\theta}_a$ also consider the shared context history $B_a$, the shared\\\hspace{1cm}reward history $b_0$ and create shared coefficients $\hat{\beta}$.
		\EndFor
		\State Choose arm that maximizes the expected reward $a_t = \argmax_{a \in A_t} p_{t,a}$, observe\\\hspace{1cm}reward $r_t$.
		\State Update the context history $A_{a_t}$ and reward history $b_{a_t}$
		\State [Hybrid] Update shared context history $B_a$ and reward history $b_0$.
		\EndFor
	\end{algorithmic}
\end{algorithm}
In the multi armed bandit problem, an agent has to make repetitive choices in a limited time frame between various actions, with each having a fixed but unknown reward probability distribution, as to maximizing the expected gain from the rewards received in each round.
As the time frame, or any other resource, is limited, the agent must constantly balance between the exploitation of promising actions and the exploration of those actions, of whose expected reward it has no good estimation yet~\cite{suttonReinforcement1998}.
Various approaches to this exploration-exploitation dilemma exist.
A commonly used one is a family of algorithms called UCB (for upper confidence bounds).
The idea is to hold a confidence interval of the expected reward for each possible action and always choose the action with the highest upper confidence bound~\cite{auerBandit2002}.

This basic algorithm, which apart from the saved intervals is stateless, can be extended to include environmental information in the so called contextual bandits.
In this work we examined one particular implementation of that algorithm family called LinUCB as first proposed by Li et al.~\cite{liContextualBandits2010}.
Lets assume an agent is put into an environment, in which it has to decide between various actions (e.g. moving a piece in a game of chess) in discrete timesteps $t = t_0, t_1 \dots$.
In each timestep, the agent perceives the environment (e.g. the positions of the pieces on a chess board) before making a decision.
In LinUCB, this perception at time~$t$ is encoded as a context vector $x_t \in \mathcal{R}^d$.
The action at time $t_i$ is chosen by computing a ridge (linear) regression between the already observed context vectors $x_{t,a}$ and the resulting reward value $r_t$ for each timestep $t = t_0 \dots t_{i-1}$ and each action $a$, thus yielding the expected reward for choosing each action in the current situation.
Exploration is promoted by adding the standard deviation to the expected reward.

Due to the complexity of the algorithm it cannot be explained in full, therefore we will only present a brief pseudo-code in Algo. \ref{linucbdisjoint} at this point.
For details please refer to the original paper~\cite{liContextualBandits2010}.
Li et al. considered two versions of the algorithm, one where each action/arm only considers previous contexts in which this action was chosen (called disjoint (context) model) and another version in which the arms have additional shared context informations (called hybrid model).
The lines marked with [Hybrid] are only considered with the hybrid model.

\subsection{Q-Learning}
\begin{algorithm}[t]
	\caption{Q-Learning based on Sutton and Barto~\cite[S.~149]{suttonReinforcement1998}.}
	\label{algqlearning}
	\begin{algorithmic}[1]
		\State Initialize $Q(s,a)$ arbitrary
		\ForAll{Episode}
		\State Initialize $s \gets s_0$
		\While{State $s$ is not terminal}
		\State Choose $a$ from $s$ policy regarding Q, e.g. $\epsilon$-greedy
		\State Execute action $a$, observe reward and state $r_{t+1},s_{t+1}$
		\State $Q(s_t,a_t) \gets Q(s_t,a_t) + \alpha[r_{t+1} + \gamma \text{ max}_a Q(s_{t+1}, a)-Q(s_t,a_t)]$
		\State $s \gets s_{t+1}$
		\EndWhile
		\EndFor
	\end{algorithmic}
\end{algorithm}
The Q-Learning algorithm computes a mapping of action-state-pairs to a real number, which represents the value for the agent of taking the action in the given state.
\begin{equation}
Q:S \times A \rightarrow \mathbb{R}
\end{equation}
Initially all Q-values are fixed to a problem-specific starting value.
Every time the agent receives a reward $r_{t+1}$ for doing action $a$ in the current state $s$ the value for this action-state-pair is updated:
\begin{equation}
Q(s_t,a_t) \leftarrow Q(s_t,a_t) + \alpha\left[r_{t+1} + \gamma \max_a Q(s_{t+1}, a)-Q(s_t,a_t)\right]\phantom{a}\alpha,\gamma \in [0,1] 
\end{equation}
$\alpha$ is the learning rate, with which new information is incorporated.
$\gamma$ is the factor for discounted rewards.
On the basis of the Q-values an agent can decide the approximate 'profitability' of choosing a certain action in a certain state.
For a complete algorithm this mechanism needs to be extended by an action choosing policy.
In this work we used a policy called $\epsilon$-greedy.
This policy accepts a probability parameter $\epsilon \in $~(0,1].
Each time the agent needs to make a decision this policy will choose the optimal action (according to the Q-values) with a (high) probability of $(1-\epsilon)$ or, with probability $\epsilon$, randomly uniform one of the non-optimal actions.
Randomly choosing a non-optimal action from time to time promotes the exploration as previously mentioned~\cite{suttonReinforcement1998}.

\section{Methods}
This section briefly describes the frameworks, the input data and the ambient process of the experiments conducted in this work.

\begin{figure}[htbp]
	\begin{center}
	\includegraphics[width=0.85\textwidth]{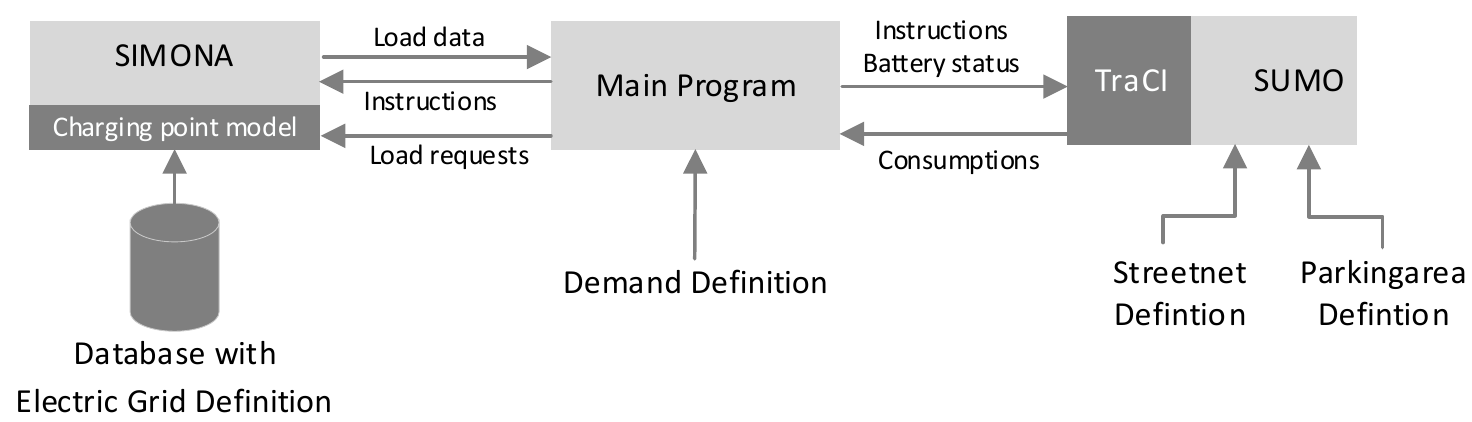}	
	\caption{Overview over the various tools used for the experiments.}
	\label{figarchitecture}
	\end{center}
\end{figure}

\subsubsection{SUMO}\hspace{-7pt}, short for Simulation of Urban MObility, is an open source software package for the simulation of traffic flows~\cite{krajzewiczSumo2016}.
It is microscopic (simulating each individual vehicle), inter- and multimodal (supporting multiple types of vehicles and pedestrians), space-continuous and time-discrete.
The tool has been chosen as it allows to simulate energy/fuel consumptions and external online intervention into vehicle behaviors using an socket API.
The simulation requires various input definitions for the street network and vehicle's mobility demands, which define where and when vehicles enter and leave the environment.
We used the tool to a) accurately measure energy consumptions of electric vehicles and therefore the additional demand for the electric grid b) to determine where and when vehicles park near charging stations.

\subsubsection{SIMONA} is a multi agent simulation tool for electric distribution grids~\cite{kaysSimona2015}.
It integrates various heterogeneous grid elements which can react on the observed power system state.
The main purpose of SIMONA is the calculation of time series under varying future scenarios and the effect of intelligent grid elements for use in grid expansion planning.
Due to the agent structure the elements can be individually parameterized and actively communicate with each other, which is beneficial for considering intelligent control elements like the one examined in this work.
Like SUMO, the simulation in SIMONA acts in discrete time steps, simulating loads, generators and other grid elements bottom up in the context of weather and other non\-electrical data to determine the power system's state.
This state includes the load flow in each time step from which the strain/loading put on each grid element can be derived.

\subsubsection{Input data}
The tool chain requires various input definitions as depicted in fig.\,\ref{figarchitecture}.
Our goal was to create a scenario that represents typical vehicle flows in a city.
Fortunately several authors have already created realistic street net and demand definitions of several European cities.
We chose the Luxembourg SUMO Traffic dataset created by Codec\'a et al.~\cite{codecaLust2017} as the city fits our needs and the representation is of high quality.
The authors showed that the demands included in the dataset realistically recreate the actual traffic in Luxembourg.
For usage in the experiments however two issues arise.
The dataset contains individual trips between two locations, each having a starting time and a random id, without the possibility to identify which trips belong to the same vehicle.
Each trip generates a vehicle in SUMO that is spawned upon the starting time and removed from the simulation as soon as it reaches it's destination.
Furthermore, the dataset contains no parking areas.

\begin{figure}[htbp]
	\begin{center}
		\includegraphics[width=0.60\textwidth]{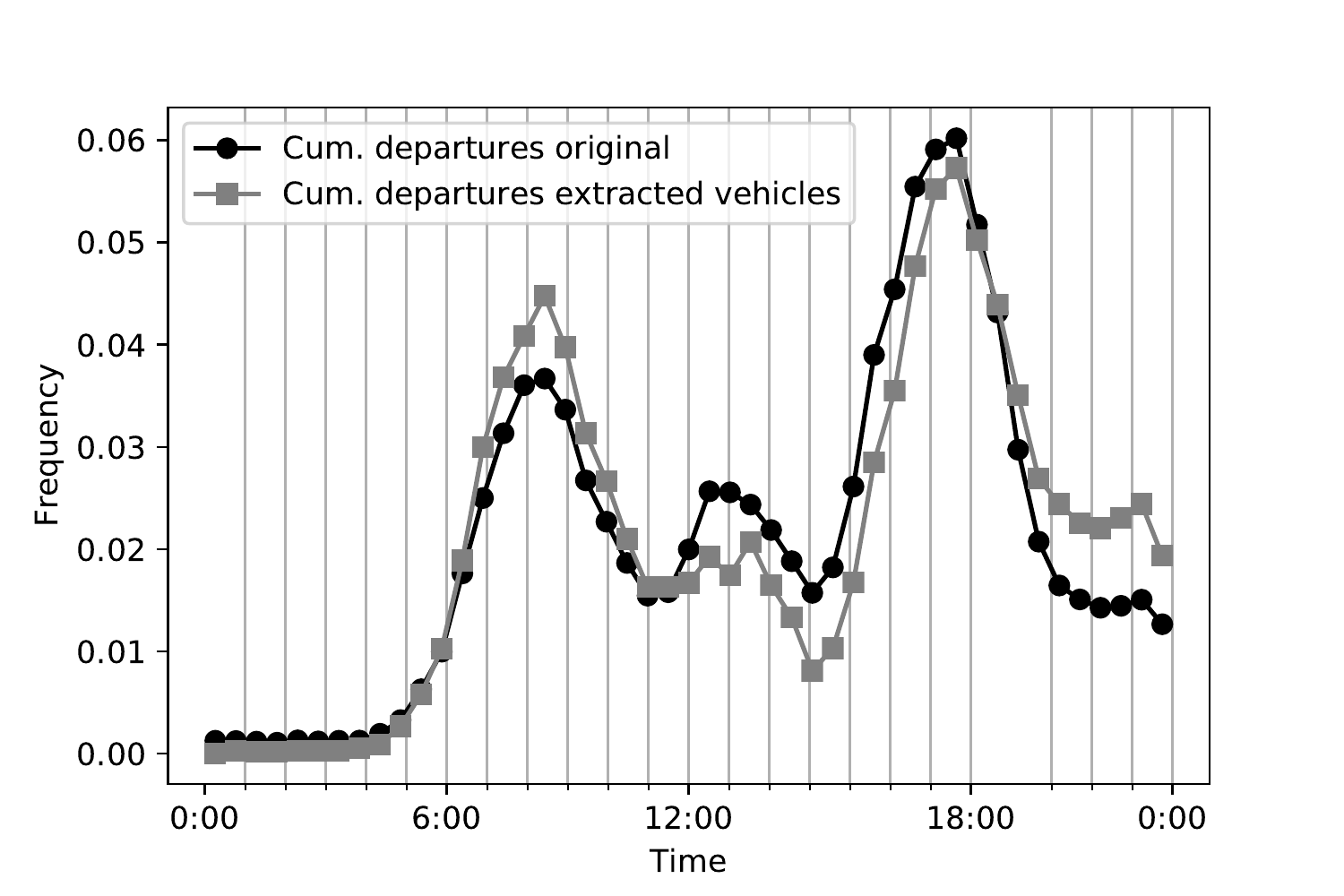}
		\caption{Cumulated normalized departures per half-hour in the original Luxembourg dataset and the extracted tours for electric vehicles.}
		\label{figdepartures}
	\end{center}
\end{figure}

As the continuity of individual vehicles, with preserving their battery state, is vital to this work we undertook measures to identify trips belonging to the same logical vehicle.
We aimed to find cycles of trips of length 2 - 4, with each trip starting on the same net edge the last trip ended on and them being ordered by the starting time, meaning the last trip of the cycle had to start last on a given day.
We discovered these circles by building a directed graph with each trip being a node.
For each pair of nodes $(f,g)$, the graph contains an edge $(f \rightarrow g)$ if the starting edge of $g$ matches the ending edge of $f$ and the depart time of $g$ is later than that of $f$.
Inside this graph, the trip circles could be found using a depth-first-search.
Using this method, a total of 13934 trip circles / vehicles have been identified, which use 27567 single trips (12.8\% of all original trips).
Fig.\,\ref{figdepartures} shows the normalized distributions of the departures in the original dataset and in the extracted tours for the electric vehicles.

The energy and charging models for the electric vehicles are based on the work of Kurczveil, L\'opez and Schnieder~\cite{kurczveilEnergy2013}, who developed an integration of electric vehicles and inductive charging stations into SUMO.
We adopted their computation system for conductive charging.
The model has been parameterized by using the values (e.g. vehicle mass, front surface area, maximum battery capacity i.a.) of the most popular pure electric vehicle (by stock) in Germany with a 22 kWh battery~\cite{kbaBestand2018}.
The manufacturer states a realistic range of about 107 km in winter.
We conducted a verification experiment in which vehicles drove random routes trough the simulated city until their battery was completely depleted.
Using the energy model of Kurczveil et al., we measured an average range of 104.0 km with a standard deviation of 6.0 km, leading us to the assumption that the model and its parameterization are sufficiently realistic.

For the simulation of the electric nets SIMONA requires a net definition, containing the various elements and their parameters which are to be simulated.
The Cigr\'e Task Force C6.04.02 created datasets of representative nets with typical topologies found in Europe and North America~\cite{cigreTaskforce2014}.
These datasets are suitable for research purposes.
The document contains three different topologies for European low voltage grids, one each for residential, commercial and industrial focuses.
To determine the count, position and type of the nets we used openly available data from OpenStreetMap (OSM), especially the coordinates of substations and land uses.
Each substation position in OSM has been used as the base coordinate of one grid.
In the next step, each of the previously extracted parking areas were assigned to the nearest grid (by euclidean distance).
After that the grid was rotated and stretched as to minimize the average distance between the net nodes and the respective parking areas.
The grid type was determined by considering the closest land use definition which had to be either residential, commercial or industrial in OSM.
This process resulted in the creation of 60 grids.

\subsubsection{Process}
On startup, the program initializes SUMO and SIMONA with the parameters stated before and creates the simulated vehicles and parking areas / charging stations.
The charging stations are associated with their respective grid nodes in SIMONA.
The simulation acts in discrete time steps of 1 second each.
In each time step, the vehicles are updated first, updating their position of they are underway or checking whether the next departure time has been reached if they are parked.
When reaching a charging station with free capacity the charging station's decision agent is updated with current data (the loading point's relative load, the current load of the substation belonging to the loading point and the load of up to 5 neighboring substations, the current time and the vehicle's current battery state of charge) and enabled to take an experiment-specific action (e.g. changing the station's offered charging price or the offered maximum power).
After that the vehicle agent can decide whether it starts the charging process.
The charging process cannot be interrupted once started except when the vehicle is leaving the parking area.
Additionally, the process can only be started when the vehicle arrives at the station.

When a vehicle leaves a parking area it receives information about the current status/offers of charging stations in walking range of their intended target.
The vehicle agent can use this information to divert from their original target and for example get a cheaper charging price. The loads of all charging points are averaged over 5 minutes each and synchronized every 300 time steps with the SIMONA framework.

\section{Experiments}
To evaluate the learning algorithms we conducted experiments in which the algorithms were used to control the behaviors of charging stations.
In the following section we define a game that is being played by the charging station agents.

\subsection{Game description}
Using the definition of Russell and Norvig~\cite{russelAI2010}, the game consists of a sequence of states which can be defined via six components.
\begin{itemize}
	\item $S_0$ (initial state): The simulation starts on the 03. January 2011 (Monday) at 00:00. All vehicles start at the starting edge of their first planned (trip) with a 100\% charged battery. Loading prices are initialized to 0.25\euro / kWh (where applicable). The initial state of the electric grid results from the first load flow calculation in SIMONA.
	\item ACTIONS$(s)$ (listing of possible actions that the agent can take in state $s$): We examined two different action models and two target variables (changing the charging price and changing the offered charging power). The variants are
	\begin{itemize}
		\item Variant A:
		\begin{itemize}
			\item The agent can take three different actions
			\begin{itemize}
				\item Increase charging price
				\item Decrease charging price
				\item Keep charging price
			\end{itemize}
			\item The action ''increase charging price'' is valid, when the price has not been decreased in the current time step yet and a defined maximum price has not been reached yet. ''Decrease charging price'' is analogous.
			\item The action ''keep charging price'' is always valid.
			\item This variant is only compatible with the 'Price' target variable.
		\end{itemize}
		\item Variant B
		\begin{itemize}
			\item The agent can take five different actions:
			\begin{itemize}
				\item Set charging price/power to 10\%, 25\%, 50\%, 75\%, 100\%.
			\end{itemize}
			\item All actions are always valid.
		\end{itemize}
	\end{itemize}
	\item PLAYER$(s)$ (determines which player/agent is choosing the next action): In every time step all agents belonging to charging stations of parking areas on which a vehicle arrived in that time step need to decide on an action. If this happens for multiple agents at the same time the order is chosen randomly.
	\item RESULT$(s,a)$ (state transition model of action $a$ in state $s$): Every action causes a change of the respective parameter (charging price or maximum power). The exact state transitions are determined by the simulation environment. The follow-up state of the last issued action in time step $t$ arises as soon as the next action is required in a time step $(t+x)$. 
	\item TERMINAL-TEST$(s)$ (tests whether the state $s$ is a terminal state, marking the end of the simulation run): The terminal state is reached after 864000 time steps (translating to 240 hours of simulated time).
	\item UTILITY$(s,p)$ (utility function of player $p$ in state $s$): We examined two different utility functions. Let $\bar{l_t} \in [0,1]$ be the average load and $(max\phantom{.}l_t) \in [0,1]$ the maximum load of the respective substation. Let $\bar{c}$ [\euro / kWh] be the average charging price. Let $m$ [\euro] be the average income (charging price~$\ast$~charged energy) of the charging pole operator. Let $\gamma \in \mathcal{R}$ be a balancing factor which constitutes a simulation parameter. We defined the utility functions as follows:
	\begin{equation}
	(\bar{l_t} \cdot -1) + ((\text{max } l_t) \cdot -1) + (\gamma \cdot \bar{c} \cdot -1)\hspace{1cm}\text{Variant 'Price'}
	\label{equtility1}
	\end{equation}
	\begin{equation}
	(\bar{l_t} \cdot -1) + ((\text{max } l_t) \cdot -1) + (\gamma \cdot m)\hspace{1.22cm}\text{Variant 'Income'}
	\label{equtility2}
	\end{equation}
	The average and maximum loads affect the reward negatively.
	The motivation of the first variant was to reduce the price as much as possible to attract customers without overloading the grid elements.
	In the second variant the pricing has been replaced by the specifically rendered service (in form of the generated income), which the charging station operator aims to maximize.
	In both cases there is a conflict of interest between the charging station operator (aiming to generate income through high power throughput) and the grid operator (aiming to prevent overloadings), which the agent both accounts for.
\end{itemize}

Formally this game definition results in a multi-objective optimization problem for each net, which involves the minimization of $\bar{l_t}$ and $(\text{max } l_t)$ and the minimization of $\bar{c}$ or maximization of $m$ respectively, with the target variables being dependent on the set of taken actions of all charging station agents.
For this definition we assume the maximization of $m$ to be equivalent to the minimization of $-m$ to reach a consistent notation.
We formalize the actions taken by each agent as integer numbers, and each agent's solution to the game as a vector of $k$ possible actions taken in $t$ possible time steps.
\begin{align}
&\min(\bar{l_t}(x), \text{max }l_t(x), \bar{c}, -m)\\\nonumber
&\hspace{0.3em}\text{s.t.}\phantom{a}x \in X\hspace{0.2cm}\text{by }X \in \mathbb{Z}^{k,t}
\end{align}
The complete solution to the game would consist of $n$ agent's solutions, with $n$ being the number of charging station agents participating.

\subsection{Strategy profiles}
We examined multiple strategy profiles which are to be described right now.
After this description the profiles will be referenced by their bold shorthand.
The profiles are determined by the following charging station agent behaviors and the utility functions (\ref{equtility1}) and (\ref{equtility2}) defined in the last section.
\begin{description}
	\item[ConstantLoading] The charging point will never change its offered price/power. The offered power is always 100\% of the maximum value.
	\item[WorkloadProportional] The charging point will change its price/power in proportion to the load of the respective substation.
	\item[Random] The price/power is determined randomly between two set thresholds.
	\item[LinUCB\_Disjunct] The agent uses a LinUCB-bandit algorithm with disjunct contexts to determine the price/power.
	\item[LinUCB\_Hybrid] Like LinUCB\_Disjunct, but with hybrid contexts.
	\item[QLearning] The agent uses a Q-Learning-algorithm to determine the price/power.
\end{description}
The behavior of the vehicles is determined by their charging and diversion behavior.
For the charging behavior we examined these two main variants:
\begin{description}
	\item[AlwaysLoad] The vehicle always starts the charging process, if it has the possibility to.
	\item[PriceAware] The vehicle holds a history of the last seen charging prices and only starts the charging process if the following condition holds. Let $c_\text{akt}$ [\euro~/~kWh] be the currently offered price. Let $C$ be the saved price history. Let $b_\text{SoC}$~[\%] be the battery state of charge.
	\begin{equation}\frac{b_\text{SoC}}{100} \leq \frac{|c \in C : c \leq c_\text{akt}|}{|C|}\end{equation}
\end{description}
We also examined variants in which the vehicles only charge 'at home' (that is, at the first charging station of the day).
Besides of these behaviors the vehicles will always charge if the battery state of charge falls below 20\% to meet basic comfort requirements.
Two different diversion behaviors have been considered:
\begin{description}
	\item[DoNotDivert] The vehicles do not change targets.
	\item[DivertToCheapest / DivertToHighestPower] The vehicle can change the target to an alternative charging station in walking distance to the desired target edge.
\end{description}
The DivertToCheapest behavior was always used when the price was the variable controlled by the charging station.
The DoNotDivert behavior was used with the ConstantLoading behavior do determine the load in the uncontrolled case.
In other experiments the DivertToHighestPower behavior was used.

For simplicity, all parking areas were equipped with charging stations with a fixed maximum charging power of 11 kW per space.
There was no distinction between private and public charging points in the simulation.
We conducted two series of experiments.
In the first one we aimed to determine the effects of the uncontrolled charging to the simulated electrical grid.
The simulation was run once without electric vehicles (to determine the base case) and once with the profiles ConstantLoading/AlwaysLoad/DoNotDivert.
The second series was run to determine the effects of the learning algorithms with varying configurations.

\section{Results}
\begin{figure}[htbp]
	\begin{center}
	\includegraphics[width=0.75\textwidth]{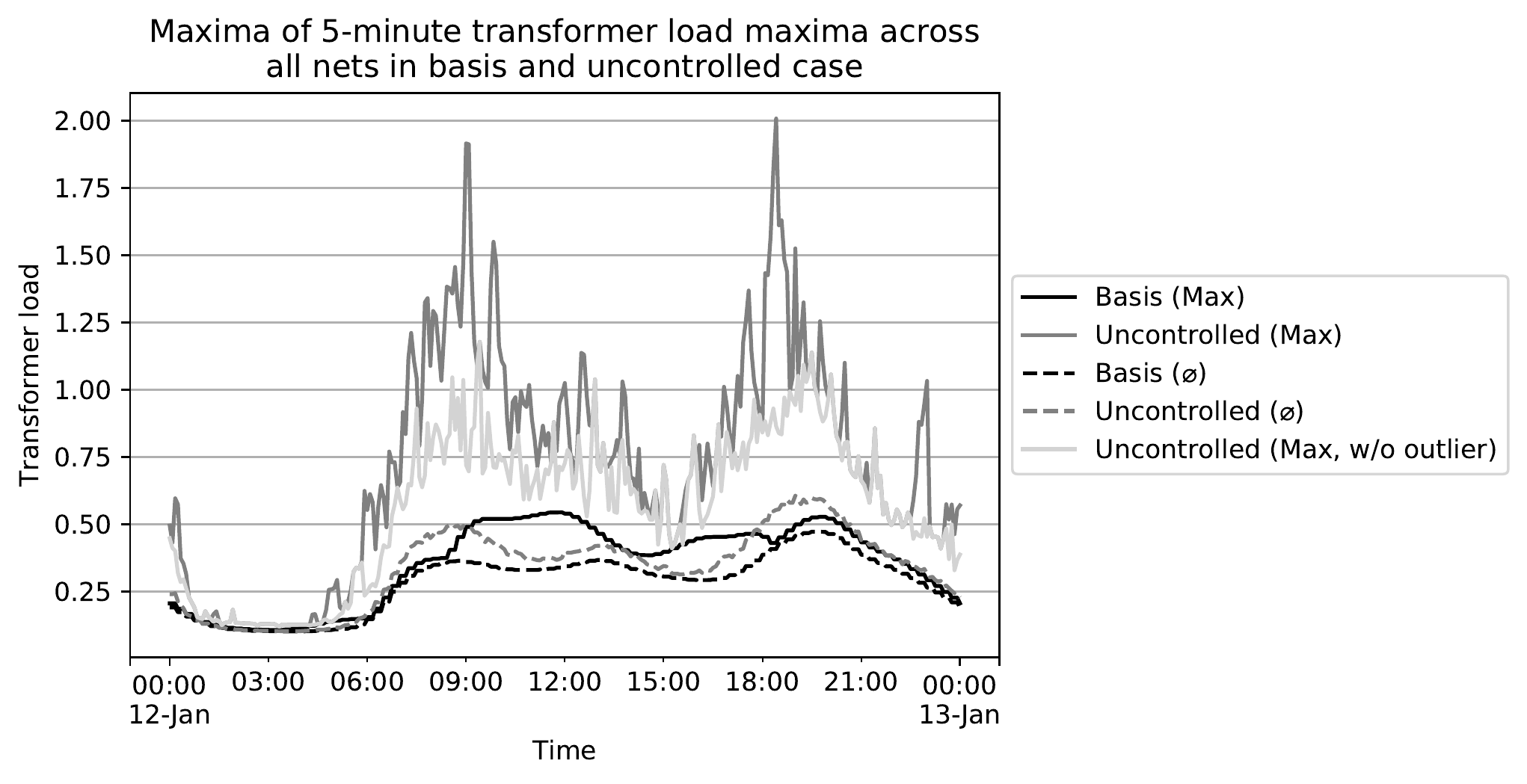}
	\caption{Maxima and average values over all 5-minute transformer load maxima (i.e.: the maximum load value of every 5 minute step is taken for each transformer and of these 60 values the maximum / average is taken) in the base case (without electric vehicles) and the uncontrolled charging case.}
	\label{figbaseuncontrolledbase}
	\end{center}
\end{figure}
\begin{figure}[htbp]
	\begin{center}
		\includegraphics[width=0.70\textwidth]{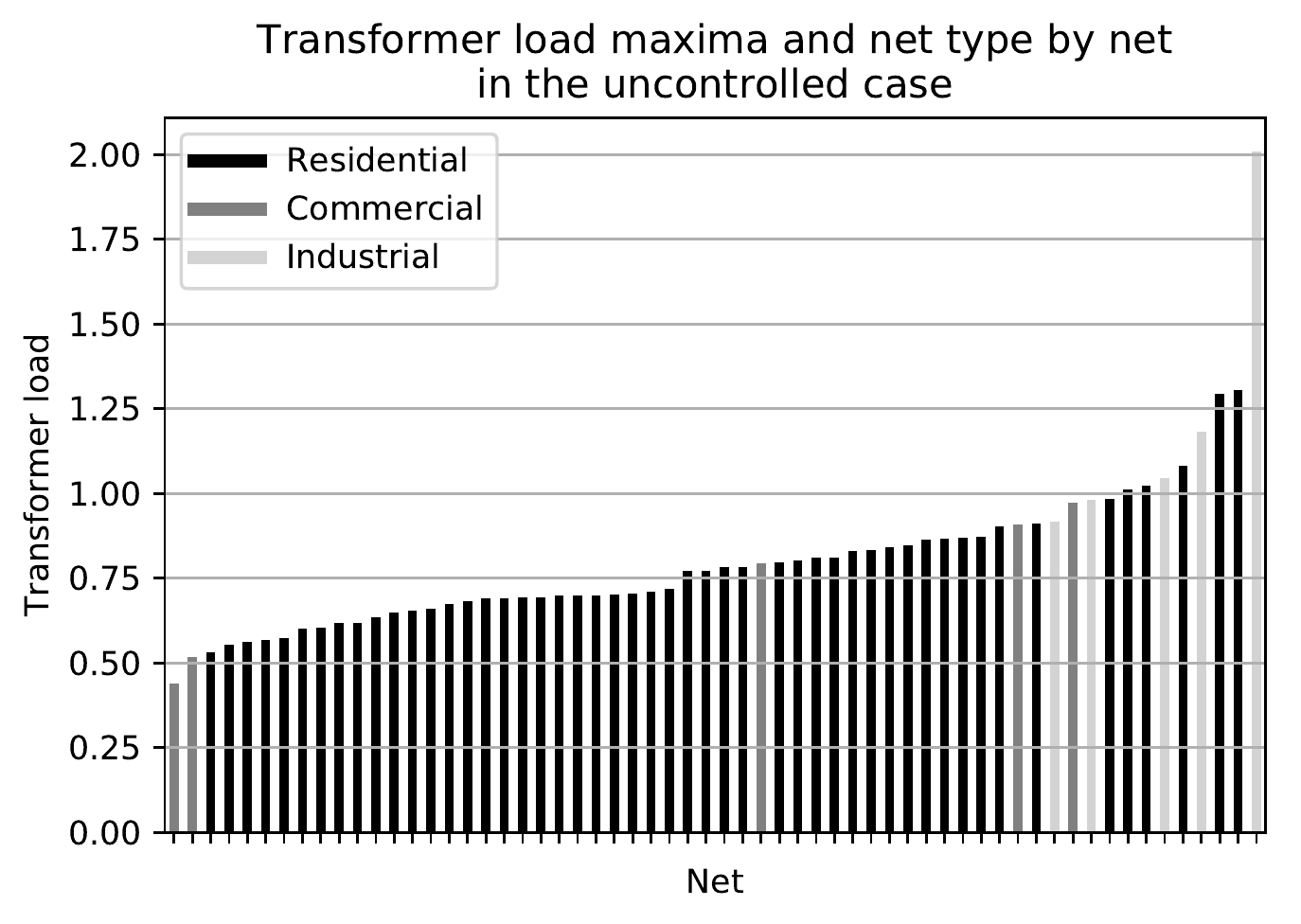}
		\caption{Transformer load maxima over one day and net type by net in the uncontrolled case.}
		\label{figloadmaximanettype}
	\end{center}
\end{figure}
For the first experiment series, our thesis was that the uncontrolled charging of electric vehicles will lead to problems in at least some grid elements.
As fig.\,\ref{figbaseuncontrolledbase} shows, that the maximum transformer load over the course of one day come to 54.4\% of the transformer's rated power in the base case but rises up to a peak of over 200\% in the uncontrolled charging case.
Note that this value is dominated by one outlier net which receives an exceptional level of traffic, however transformer loads between 101.2\% and 130.5\% were measured in 7 more nets, which means 8 / 60 subnets registered at least one overload situation, as can be seen in fig.\,\ref{figloadmaximanettype}.
There were no overloads in other net elements like power lines or significant voltage deviations.

In the second experiment series we took a deeper look into the effects of the various control algorithms for charging stations on the grid.
Our thesis were that the learning algorithms improved their behavior (measured by the received reward values) over time, that the bandit algorithm reduced the negative effects of the vehicle charging loads on the grid and that the learning algorithms all in all perform better than the simpler ones.
\begin{figure}[htbp]
	\begin{center}
	\includegraphics[width=0.80\textwidth]{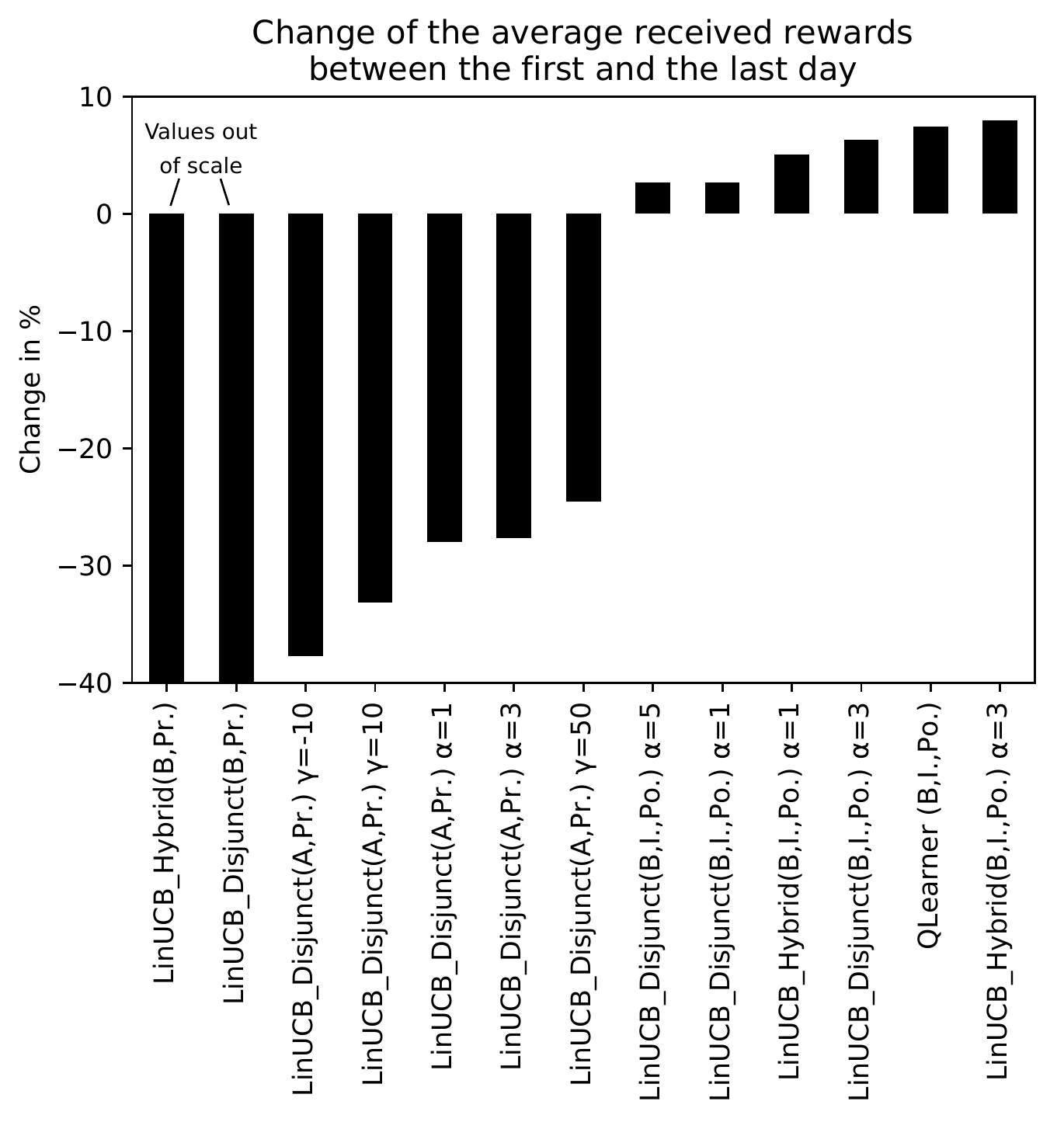}
	\includegraphics[width=0.80\textwidth]{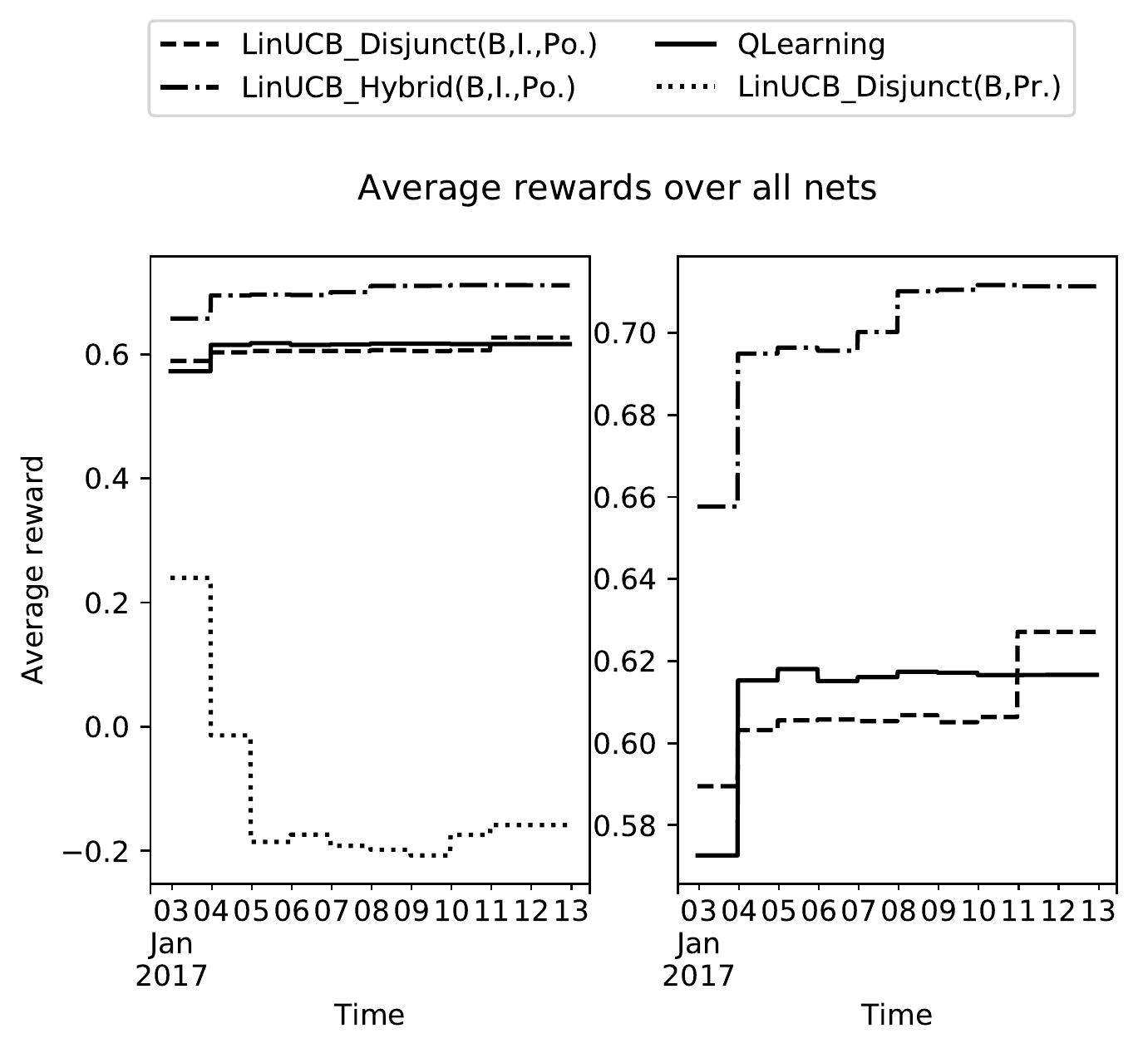}
	\caption{Top: Change of the average received reward between the first and
	last day. Bottom: Average received rewards between the first and last day for selected algorithm variants.}
	\label{figchangeofaveragerewards}
	\end{center}
\end{figure}
Fig.\,\ref{figchangeofaveragerewards} top shows the absolute change between the averaged received rewards of the first and last day over all charging point agents for various selected profiles.
The values in the parentheses state the used action variant (A or B), the utility function (I. = Income, P. = Price) and the target variable (Po. = Power, Pr. = Price).
It can be seen that some variants of LinUCB and QLearning develop positively (in the sense of received rewards), which indicates that the learning target is being approached.
One must note that many parameter combinations did not perform very well.
The agent which used the agent model variant A or the variant B with the price as a controlling variable rarely converged towards a meaningful result.
Also, in some configurations the choice of the balancing factor $\gamma$ or the LinUCB-parameter $\alpha$ had a significant impact on the overall performance even for small changes.
Fig.\,\ref{figchangeofaveragerewards} bottom shows the development of the received rewards for selected scenarios.
The values have been normalized to [0,1] for better comparability.
It is noticeable that the general trend of the development can usually be seen after just 1 day of simulation.

\begin{figure}[htbp]
	\begin{center}
		\includegraphics[width=0.69\textwidth]{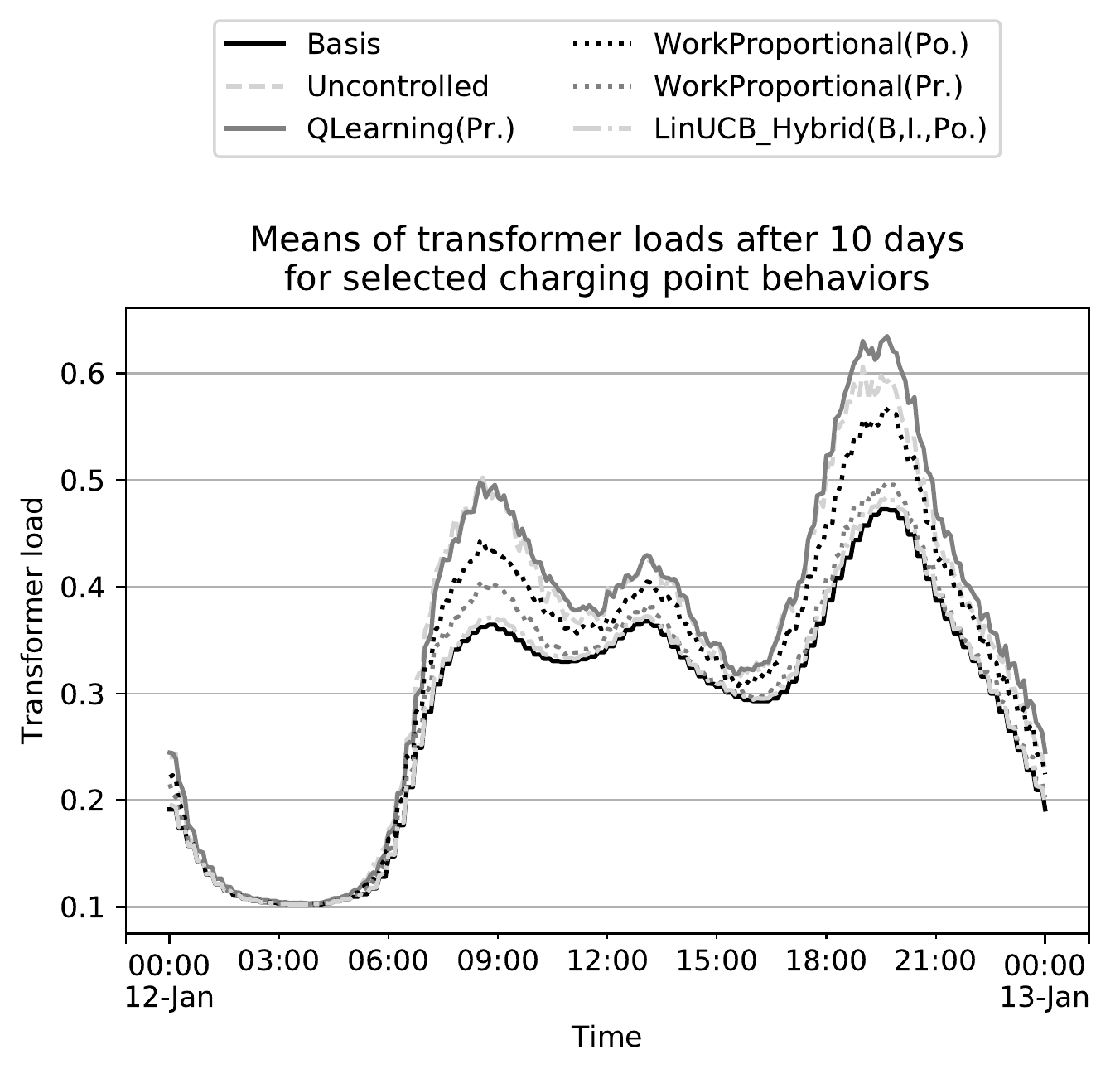}
		\caption{Mean transformer loads over all nets after 10 days for selected charging point behaviors.}
		\label{figaverageloads}
	\end{center}
\end{figure}
In the next step we examined the transformer loads after 10 days of simulation using the various charging point behaviors and their respective control algorithms (fig.\,\ref{figaverageloads}).
Note that for a better overview the plot only shows the best performing algorithm variant for each category.
The random-choosing profile has been left out as it did not perform very well.
The mean transformer load could be reduced by up to 12\% in comparison to the uncontrolled case, the maximum peak load could be reduced from 201\% to 70\%.
The Q-Learning-algorithm, while making some progress reward-wise (fig.\,\ref{figchangeofaveragerewards} bottom), did not perform very well in this scenario.
The training time possibly was not long enough for it to converge towards a profitable solution.
While the best-performing algorithm was a variant of LinUCB using action variant B, the ''Income'' utility function (2) and a power target variable, a variant of the WorkProportional-strategy performed almost as good.
In contrast to the bandit algorithms, the simple price-centered WorkProportional-strategy performed comparably well.

\FloatBarrier
\section{Conclusion}
With the proliferation of electric vehicles, the electrical distribution grids
are more prone to overloads. In this paper we provided a literature survey on
countermeasures to control charging of electric vehicles such that overloads
are prevented. After a brief introduction of the fundamentals, we modeled and
studied an intelligent pricing
mechanism based on a reinforcement learning problem. As context information
is crucial in our setting, we tested in particular contextual bandit learning
algorithms to provide incentives for distributing
charging load and prevent network failures. 

The presented algorithms were implemented and combine the
microscopic mobility simulator SUMO with the electrical network simulator
SIMONA. The simulation framework thus produces reliable electrical distribution load values. 

Our extensive experiments are carefully conducted under realistic conditions and reveal that conditional bandit learning outperforms context-free reinforcement learning algorithms and our approach is suitable for the given problem.
While we found that the used bandit algorithms were indeed able to reduce the problematic effects on the grid considerably, we also noticed that some of the tested variants did not perform very well in the simulation environment.
From this we conclude that, if the algorithm were to be implemented productively, considerable work would need to be invested into the correct parametrization.
After this has been accomplished a learning algorithm should be able to be rapidly implemented in various target environments.

Due to the rising popularity of electric mobility the charging stations will become a vital part of future mobility and transportation considerations. As reinforcement learning algorithms can be adapted rapidly to include new information we assume these to be suitable as part of a holistic traffic control scenario.

In future works, the decision model of the vehicle (passengers) could be expanded.
As the emphasis of this work lied on the charging stations a simple vehicle's model was chosen.
A more complex model that considers the planned daily routine, retention times or socioeconomic factors could lead to a more diverse task and thus promote the advantages of self-adapting charging stations even better.
Future works should also consider differences between private and public charging points: A charging device owned by the same person as the vehicle would pursue other goals than a profit-oriented public charging station.
A private charging device could potentially be a useful actor in a holistic smart home environment which can also include a privately owned photovoltaic system.
Lastly the agent system was designed with an emphasis on independence of single charging points.
Another interesting approach would be to test a mechanism design in which the agents act towards a common target and are rewarded/graded as an ensemble and not individually.
This would be realistically possible for operators owning multiple charging stations, as they probably run a centralized controlling platform.

\subsection*{Acknowledgements}

Part of the work on this paper has been supported by Deutsche
Forschungsgemeinschaft (DFG) within the Collaborative Research Center SFB 876
''Providing Information by Resource-Constrained Analysis'', project B4.
Thomas Liebig received funding by the European Union through the Horizon 2020
Programme under grant agreement number 688380 ''VaVeL: variety, Veracity,
VaLue: Handling the Multiplicity of Urban Sensors''.

This work contains results from the master's thesis of Christian R\"omer titled ''Ladesteuerung von Elektrofahrzeugen mit kontextsensitiven Banditen unter Ber\"ucksichtigung des elektrischen Verteilnetzes'' at the TU Dortmund University.

\bibliography{references}

\end{document}